\documentclass[11pt]{article}
\usepackage[T1]{fontenc}
\usepackage[latin1]{inputenc}
\usepackage{graphics}
\usepackage{setspace}
\makeatletter

\newcommand{\LyX}{L\kern-.1667em\lower.25em\hbox{Y}\kern-.125emX\spacefactor1000}

\makeatother

\begin{document}

\title{Protein Unfolding and the Diffusion Collision Model}

\author{Chris Beck\protect\( ^{a}\protect \) and Xavier Siemens\protect\( ^{b}\protect \)}

\maketitle
{\centering \emph{\( ^{a} \)Molecular Modeling Laboratory and \( ^{b} \)Institute
of Cosmology}\par}

{\centering \emph{Department of Physics and Astronomy}\par}

{\centering \emph{Tufts University}\par}

{\centering \emph{Medford, MA 02155}\par}

\begin{abstract}
In the diffusion-collision model, the unfolding or backward rates are given
by the likelihood of secondary structural cluster dissociation. In this work,
we introduce a backward rate calculation modeled from a Kramers-type thermal
tunneling through a barrier, which represents the free energy potential well
for buried hydrophobic residues. Our results are in good agreement with currently
accepted values and the approach suggests a link between the diffusion-collision
and folding funnel models of protein folding.
\end{abstract}

\section*{I. Introduction}

In the diffusion-collision model (DCM) of protein folding (Bashford et al 1988,
Karplus and Weaver, 1976, 1979, 1994) the protein is modeled using a collection
of spheres connected by flexible strings. The spheres represent the secondary
structural elements such as \( \alpha  \)-helices or \( \beta  \)-sheets,
called microdomains, that make up the protein. 

The folding process from a completely unfolded protein to the the final native
state is accomplished via diffusion through the solvent, collision, and finally
coalescence of the microdomains. The state of the protein is defined by the
number of pairings (hydrophobic interactions) between the microdomains that
are present at a given time \( t \). The rate equations for transitions between
these states can be written as
\begin{equation}
\label{rateeqn}
\frac{d{\bf P}(t)}{dt}={\hat{K}}{\bf P}(t)
\end{equation}
where \( {\bf P}(t) \) is the vector of states and \( {\hat{K}} \) is a matrix
containing the transition rates between the different states. 

As an example, consider a simple monomeric protein containing two \( \alpha  \)-helices
A and B joined by a connecting string. This gives us a one-pair/two-state system.
We will call state \( 1 \) the state when the two microdomains are not in hydrophobic
contact, and state \( 2 \) when they are hydrophobically docked. In this case
(\ref{rateeqn}) can easily be written out explicitly

{\centering 
\begin{equation}
\label{couplerate}
\begin{array}{c}
\dot{P}_{1}(t)=-k_{1\rightarrow 2}P_{1}(t)+k_{2\rightarrow 1}P_{2}(t)\\
\dot{P}_{2}(t)=k_{1\rightarrow 2}P_{1}(t)-k_{2\rightarrow 1}P_{2}(t)
\end{array}
\end{equation}
\par}

or in matrix form as

\begin{equation}
\label{couplerate2}
\frac{d}{dt}\left( \begin{array}{c}
P_{1}(t)\\
P_{2}(t)
\end{array}\right) =\left[ \begin{array}{cc}
-k_{1\rightarrow 2} & k_{2\rightarrow 1}\\
k_{1\rightarrow 2} & -k_{2\rightarrow 1}
\end{array}\right] \left( \begin{array}{c}
P_{1}(t)\\
P_{2}(t)
\end{array}\right) .
\end{equation}

A more complicated protein having, say, \( n \) microdomains would involve
\( p=n(n-1)/2 \) pairings, \( 2^{p} \) states \( P_{i}(t) \) and a \( 2^{p}\times 2^{p} \)
rate matrix \( \hat{K} \). In general the calculation of the elements of the
rate matrix \( {\hat{K}} \) is somewhat involved. The forward rates are the
rates of microdomain coalescence. In the diffusion-collision model the forward
rates are calculated assuming the microdomains diffuse through a solvent environment,
the space of which is limited by the length of the intervening strings and the
van der Waals radii of the microdomains. These microdomains are assumed to be
nascently formed, and their degree of formation is given by a helix-coil transition
theory (Zimm and Bragg, 1959, Lifson and Roig, 1961) as in AGADIR (Lacroix et
al, 1998, Munoz and Serrano, 1994 a,b,c, 1997) calculation in the case of \( \alpha  \)-helices,
or via a combination of theory (Munoz et al, 1998) and experiment (Dinner et
al, 1999) in the case of \( \beta  \)-sheets. As the microdomains undergo diffusion,
they occasionally collide. When this happens the microdomains coalesce with
a probability \( \gamma  \), being held together by hydrophobic interactions
in the case of \( \alpha  \)-helices, or a combination of hydrophobic and hydrogen
bond interactions in the case of \( \beta  \)-sheets. The coalescence probability
\( \gamma  \) is given by the likelihood that the microdomain is in \( \alpha  \)-helical
or \( \beta  \)-sheet form, the percentage of hydrophobic area, and the likelihood
of proper geometrical orientation upon collision.

The forward folding times in the mean first passage time approximation (Weiss,
1967, Weaver, 1979, Szabo et al 1980) are given by

\begin{equation}
\label{MFPT}
\tau _{f}=\frac{l^{2}}{D}+\frac{LV(1-\gamma )}{\gamma DA}
\end{equation}
 where \emph{\( V \)} is the diffusion volume available to the microdomain
pair, \emph{\( A \)} is the target area for collisions, \emph{\( D \)} is
the relative diffusion coefficient, \( \gamma  \) is the probability of coalescence
upon collision and \emph{\( l \)} and \emph{\( L \)} are geometrical parameters
calculated for diffusion in a spherical space. The inverse of the first passage
time-scales \( \tau _{f} \) are the forward folding rates \emph{\( k_{f} \)}
that are used in the rate matrix \( {\hat{K}} \)\emph{.} In the example given
by (\ref{couplerate}) and (\ref{couplerate2}), \( k_{f} \) is \( k_{1\rightarrow 2} \).

The microdomain pairings can also dissociate. To date, in typical diffusion-collision
model calculations, the form of the backward folding, or unfolding times \( \tau _{b} \)
used for two microdomains A and B comes from from the Van't Hoff-Arrhenius law
(Van't Hoff, 1884, Arrhenius, 1889) given by

\begin{equation}
\label{brDCM}
\tau _{b}=\nu ^{-1}e^{fA_{AB}/k_{B}T}
\end{equation}
 where \emph{\( f \)} is the free energy change per unit buried hydrophobic
area in the pairing (Chothia, 1974), \emph{\( A_{AB} \)} is the buried area
(Lee and Richards, 1971), \emph{\( k_{B} \)} is Boltzmann's constant, \emph{\( T \)}
is the temperature and \( \nu  \) is an attempt rate. 

In diffusion collision model calculations to date the attempt rate is a parameter
whose value must be set by hand, usually requiring some guess work or adjustment
to obtain the desired result (see for example Burton et. al 1998). The typical
values used lie in the range \( 1\times 10^{9}s^{-1}-1000\times 10^{9}s^{-1} \),
obtained from estimates of covalent bond oscillation frequencies (Fersht, 1999).
This procedure is not explicitly stated in most diffusion-collision model studies;
it is justified, however, when the equilibrium occupation probabilities of the
states are known. In fact, the rate constants can be calculated from such probabilities
(Chandler, 1987). In cases where the final occupation probabilities are unknown,
for instance in studies of protein mis-folding and non-native kinetic intermediates
(Beck et al., 2000), such methods are clearly not possible. 

In this work we show how the parameter \( \nu  \), or more generally the unfolding
rates, can be determined from thermal fluctuations providing a means of avoiding
the guesswork. This makes the diffusion-collision model more predictive and
enables it to be used in situations where the final occupation probabilities
are unknown.

\section*{II. Calculation of the Unfolding Rates}

In order to calculate the backward rate it is convenient to view the unfolding
process as that of diffusion within, and escape from, an effective one-dimensional
potential well. This is a good approximation if only motion perpendicular to
the hydrophobic contact surface is important in microdomain pair dissociation. 

Consider the pair of microdomains A and B connected by a string. Microdomain
coalescence and dissociation can be approximated by diffusion in a potential
like the one depicted in Figure 1. The quantity that diffuses is the separation
between the microdomains and we can think of

\begin{equation}
\label{prob}
dP(x)=\rho (x)dx
\end{equation}
 as the fraction of microdomains with a distance between them in the range \( x \)
and \( x+dx \). In a potential \( V(x) \) the probability density \( \rho (x) \)
satisfies the one dimensional Smoluchowski equation

\begin{equation}
\label{smoleqn}
\frac{\partial \rho (x,t)}{\partial t}=\frac{\partial }{\partial x}\left[ D\left( \frac{\partial \rho (x,t)}{\partial x}+\rho (x,t)\frac{\partial V(x)}{\partial x}\right) \right] .
\end{equation}

\begin{figure}
{\centering \resizebox*{5cm}{7cm}{\includegraphics{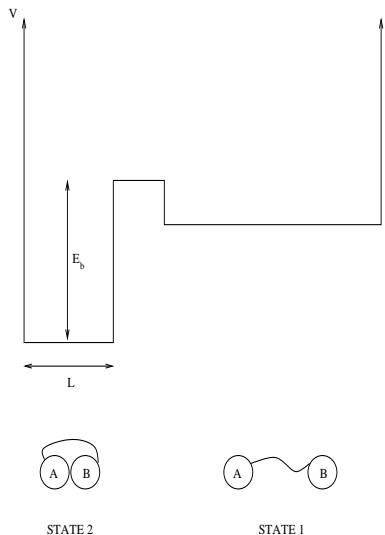}} \par}

\caption{Diffusion potential for the two microdomains A and B. The potential is infinite
on the far right because the microdomain separation cannot exceed the length
of their connecting string. On the left it is also infinite, in this case because
of the hard-core repulsion of the van der Walls contact between the microdomains.
The potential energy barrier has a height \protect\( E_{b}=fA_{AB}\protect \),
the free energy difference between paired and unpaired states with a buried
hydrophobic area \protect\( A_{AB}\protect \), and width \protect\( L\protect \),
which we take to be the diameter of a water molecule.}
\end{figure}

The outer boundary on the far right is given by an infinite potential, arising
from the two microdomains coming to the end of their connecting string. The
inner boundary is also infinite, stemming from the hard-core repulsion of the
van der Waals contact between the microdomains. The well depth \( E_{b}=fA_{AB} \)
is the free energy difference between paired and unpaired microdomains and the
well width \( L \) is set to the diameter of a water molecule. A separation
larger than \( L \) exposes the microdomain to the solvent, the free energy
savings is lost, and the microdomains separate, resulting in an escape from
the potential well.

The forward folding rates can be satisfactorily found by the mean first passage
time using the method outlined above (\ref{MFPT}). As a consequence of this
we shall forego direct analysis of the entire diffusion process and concentrate
on the escape of a bound microdomain out of the potential well in order to find
the backward folding rate. 

If we concentrate on the part of the potential where \( x<L \) in Figure 1,
we can think of the potential as being perfectly reflecting on one side and
partially permeable on the other. The permeability arises from the thermal fluctuations
in the energy which allow the microdomains to escape the bounded region, provided
they have sufficient energy, and it can be expressed using an appropriate choice
of boundary conditions. We proceed as follows.

For the inner boundary, representing van der Waals contact, we impose perfectly
reflecting boundary conditions, meaning the flux through the boundary is zero

{\centering 
\begin{equation}
\label{bc@0}
J(0,t)=D\left. \frac{d\rho (x,t)}{dx}\right| _{x=0}=0.
\end{equation}
\par}

The flux at the permeable boundary at \( x=L \) depends on the density of microdomains
at that boundary and the probability that the microdomain energy is high enough
to thermally tunnel through the boundary. The differential element of flux at
the boundary \( L \) of microdomains with relative velocity between \emph{\( v \)}
and \emph{\( v+dv \)} is given by

\begin{equation}
\label{bc@L}
dJ(L,t)=v\rho (L,t)dN(v)
\end{equation}
 where

\begin{equation}
\label{MBDist}
dN(v)=\left( \frac{\mu }{2\pi k_{B}T}\right) ^{1/2}e^{-\mu v^{2}/2k_{B}T}dv
\end{equation}
is the Maxwell-Boltzmann distribution, namely the fraction of microdomains with
relative velocities between \( v \) and \( v+dv \) and \( \mu  \) is the
reduced mass given by

\begin{equation}
\label{reducedM}
\mu =\frac{m_{A}m_{B}}{(m_{A}+m_{B})}
\end{equation}
where \( m_{A} \) and \( m_{B} \) are the masses of the two microdomains.
The distribution is normalised so that the bounds of the integral go from \( -\infty  \)
to \( \infty  \).

It is clear that in order to find the total flux through the outer boundary
at \emph{\( L \)} we must integrate over all microdomain velocities larger
than \( +\sqrt{E_{b}/2m} \) since the potential well height is \( E_{b} \),
and only microdomains with velocity higher than that can escape and thus contribute
to the flux leaving the well. This yields a flux out of state \( 2 \)

\begin{equation}
\label{totalJ@L}
J_{2}^{out}(L,t)=\rho (L,t)\left( \frac{k_{b}T}{2\pi \mu }\right) ^{1/2}e^{-E_{b}/k_{B}T}
\end{equation}

If we assume that the probability is uniformly distributed across the well,
then \( \rho (x,t)=P_{2}(t)/L \) everywhere within the well. This is a reasonable
assumption because the permeability of the well is so low. Furthermore it can
be shown that typically the damping time (or velocity autocorrelation) time
is considerably smaller than the well crossing time and the mean first passage
time. Clearly then, diffusion in the interior of the well distributes the microdomain
separations evenly among the available space and therefore 
\begin{equation}
\label{totalJ@L2}
J_{2}^{out}(L,t)=\frac{P_{2}(t)}{L}\left( \frac{k_{b}T}{2\pi \mu }\right) ^{1/2}e^{-E_{b}/k_{B}T}.
\end{equation}

The interesting result is that by using a stationary flux method (Kramers, 1940),
or equivalently setting \( P_{2}(t)=1, \) the outward flux at the boundary
in one dimension \( J^{out}(L,t) \) is the fraction of microdomain pairs crossing
the boundary at \( L \) per unit time and thus clearly identical to the unfolding
rate \( k_{b} \) for the \( AB \) microdomain pairing. Below we explain this
in more detail and show more generally how the rate equations of the diffusion-collision
model (\ref{rateeqn}, \ref{couplerate}) can be derived from the diffusion
equation.

To understand this, we remember that the Smoluchowski equation (\ref{smoleqn})
is simply a detailed statement of the equation of continuity 

{\centering 
\begin{equation}
\label{conteqn}
\frac{\partial \rho ({\bf x},t)}{\partial t}={\bf \nabla }\cdot {\bf J}({\bf x},t)
\end{equation}
\par}

The quantity in brackets in (\ref{smoleqn}) is the flux \( {\bf J}({\bf x},t) \),
which includes the ordinary \( \nabla \rho ({\bf x},t) \) term, and the second
term which takes into account external forces. Integration of both sides of
(\ref{conteqn}) over the bound part of diffusion volume, using the divergence
theorem on the R.H.S. and assuming isotropic flow into and out of the boundary
of the diffusion volume yields

{\centering 
\begin{equation}
\label{intconseqn}
\frac{dP_{2}(t)}{dt}={\bf n}\cdot {\bf J}(t)A=\{J_{2}^{in}(t)-J_{2}^{out}(t)\}A.
\end{equation}
\par}

The LHS of this equation is the rate of change of probability in the bound region.
On the RHS \( {\bf n} \) is a unit vector normal to the boundary, \( {\bf n}\cdot {\bf J}(t) \)
is the net flux which traverses the boundary surface and \( A \) is the area
of that boundary surface. We suggestively write the net flux \( {\bf n}\cdot {\bf J}(t) \)
in terms of \( J_{2}^{in}(t) \) and \( J_{2}^{out}(t) \), the probability
fluxes in and out of the diffusion volume in question. 

We see from comparison to (\ref{couplerate}) that (\ref{intconseqn}) can be
written out explicitly for our case

{\centering 
\begin{equation}
\label{indrateeqlsflux}
\{J_{2}^{in}(t)-J_{2}^{out}(t)\}A=k_{1\rightarrow 2}P_{1}(t)-k_{2\rightarrow 1}P_{2}(t).
\end{equation}
\par}

One can easily see that the positive terms on either side of (\ref{indrateeqlsflux})
are equal to each other as are the negative terms. Referring to (\ref{totalJ@L}),
this quantity is 

{\centering 
\begin{equation}
\label{fluxout}
J_{2}^{out}(t)A=k_{2\rightarrow 1}P_{2}(t)
\end{equation}
\par}

making the rate

\begin{equation}
\label{fluxovereprob}
k_{2\rightarrow 1}=\frac{J_{2}^{out}(t)A}{P_{2}(t)}.
\end{equation}

In this expression the one-dimensional ``area'' through which the probability
is flowing is simply a point and therefore \( A=1 \) so identifying \( J_{2}^{out}(t) \)
with (\ref{totalJ@L2}) we can solve for the unfolding rates in \( {\hat{K}} \),
\( k_{2\rightarrow 1} \) in (\ref{couplerate}) and (\ref{couplerate2}), by
dividing (\ref{totalJ@L2}) by \( P_{2}(t) \), or setting \( P_{2}(t)=1 \)
in (\ref{fluxovereprob}) according to the stationary flux method. We find the
backward folding rate in the one-dimensional case to be

\begin{equation}
\label{1Dbackrate}
k_{2\rightarrow 1}=\frac{1}{L}\left( \frac{k_{B}T}{2\pi \mu }\right) ^{1/2}e^{-E_{b}/k_{B}T}.
\end{equation}

The terms preceding the exponential correspond to our prediction for the Van't
Hoff-Arrhenius attempt rate \( \nu  \) in (\ref{brDCM}). As an example, the
attempt rate found for a coalesced pair of 16-residue Regan-Degrado (Regan and
Degrado 1988) helices with a combined hydrophobic area loss of \( 600 \)\AA\( ^{2} \) is
\( 64\times 10^{9}s^{-1} \). 

We believe that the most probable dissociation of a microdomain-microdomain
pairing occurs via relative motion along a vector connecting the centers-of-mass
of the two microdomains. This implies that typically the one dimensional case
is the best way to view the dissociation event and therefore that one should
use (\ref{1Dbackrate}) to go about calculating the rate. The one dimensional
case should also be sufficient if there is an ``unzipping'' of paired \( \alpha  \)-helices. 

It is possible, however, that dissociation may also include a relative rolling
motion or other motion perpendicular to the axis connecting the microdomain
pairs. In this case one needs to repeat the calculation above with a few minor
differences: The relative velocity distribution of the microdomains is still
the one dimensional Maxwell-Boltzmann distribution because the degrees of freedom
parallel to the surface through which the probability is flowing do not contribute
to escape from the well, the probability in the bound region is assumed to be
evenly distributed in a two-dimensional ``volume'', namely \( \rho (x,t)=P_{2}(t)/\pi L^{2} \)
in the two-dimensional analogue of (\ref{totalJ@L}), and the flux goes through
a two-dimensional ``area'' \( A=2\pi L \) in (\ref{fluxovereprob}). This
calculation yields the result

{\centering 
\begin{equation}
\label{2Dbackrate}
k_{2\rightarrow 1}=\frac{2}{L}\left( \frac{k_{b}T}{2\pi \mu }\right) ^{1/2}e^{-E_{b}/k_{B}T}.
\end{equation}
\par}

Due to the steric clashing of the side chains it seems rather unlikely that
dissociation would include a relative sliding motion along the axes of the microdomains.
For completeness, however, we include the three-dimensional result derivable
from analogous considerations to the ones given above

{\centering 
\begin{equation}
\label{3Dbackrate}
k_{2\rightarrow 1}=\frac{3}{L}\left( \frac{k_{b}T}{2\pi \mu }\right) ^{1/2}e^{-E_{b}/k_{B}T}.
\end{equation}
\par}

Although the three-dimensional result seems an unlikely candidate for protein
unfolding it may be relevant in the context of molten globules.

This approach succeeds in removing the free parameter \( \nu  \) from the model,
and allows us to find the backward rates from a simple energetic model based
on diffusion in a potential with appropriate boundary conditions. Moreover our
results for the one, two and three-dimensional unfolding rates have a \( \sqrt{T/\mu } \)
dependence that could be used to distinguish between this and other proposals
for the mechanism of helix-helix dissociation.

The removal of the parameter \( \nu  \) is important when considering folding
processes which do not involve the native state. In previous applications of
the diffusion-collision model, the folding kinetics from a denatured or random
coil state to the final native state were followed. In such a case, it is reasonable
to set the parameter \( \nu  \) such that the native state achieves 90 or 95
percent of the probability, because we know that the final state is attained
at the end of the folding process. In studying intermediate processes or more
importantly, non-native intermediates (Beck et al, 2000), where the occupation
probability may be completely unknown, such reasonable estimates of \( \nu  \)
are not available. In this case, elimination of \( \nu  \) as a free parameter
is crucial.

\section*{III. Concluding Remarks}

We have presented a calculation for the helix-helix dissociation rate using
a simplified potential surface, a square potential having a depth equal to the
free energy savings of hydrophobic docking and width equal to the diameter of
a water molecule. We have found the unfolding rates arising from thermal fluctuations
out of this potential well to be in good agreement with currently accepted values
of \( \nu  \). The potential itself is due to hydrophobic forces, which to
date are not well understood and for this reason the potential chosen was simple. 

The initial motivation of this work was to eliminate the free parameter \( \nu  \)
from the diffusion-collision model. In the context of this model our result
should allow us to perform kinetics simulations than were not possible before,
for instance time developments of non-native intermediates, in which the occupation
probabilities are unknown, and reasonable estimates for \( \nu  \) are not
possible. 

The results presented here also predict a \( \nu \propto \sqrt{T/\mu } \) dependence
in all cases which can be distinguished experimentally from other proposals
such as the covalent bond model where \( \nu \propto k_{B}T/\hbar  \) (Fersht,
1999). Another difference is the dependence of the unfolding rates on the states,
not only through the hydrophobic area, but also through the reduced mass \( \mu  \)
of the microdomains or groups of microdomains undergoing dissociation. This
is markedly different from typical diffusion-collision model calculations where
the attempt rate \( \nu  \) is assumed to be the same for all dissociation
events within the protein.

Along the way we have re-examined the idea that couching the problem of association
and dissociation of microdomain pairs via diffusion over a potential surface
affords us a clear and simple picture of the protein folding process. Indeed,
the rate equations of the diffusion-collision model can be derived from such
a picture. Interestingly, the potential energy surface is an element of other
models such as folding funnels (for an overview see Bryngelson et al 1995). 

Admittedly, our approximation is rather rough. The thrust of future work should
therefore be in constructing more realistic potentials including activation
energies of hydrophobic dockings. 

Generalization to more complex proteins is straightforward because every interaction
between helices or clusters of helices can be considered a two-state process
similar to the one described above; generally, the folding or unfolding of a
given protein involves several such processes. In this case the diffusion space
for each possible pairing could be constructed, and the forward and backward
rates for each transition can be found, as outlined above, to construct the
transition rate matrix \( {\hat{K}} \) in order to find the time evolution
of the state vector \( {\bf P}(t) \). This is a much more tenable proposition
than directly solving the Smoluchowski equation (\ref{smoleqn}) on such a complicated
potential surface, although it could be done in principle.

\section*{Acknowledgments}

We would like to thank Ken Olum, David Weaver, and Larry Ford for many useful
discussions. We are further grateful to David Weaver for pointing out a mistake
in our calculations and Ken Olum for helping us fix it.

\section*{References}

\textbf{Arrhenius S.} 1889. \emph{Z Phys Chem 4}:226.

\noindent \textbf{Bashford D, Cohen FE, Karplus M, Kuntz ID, Weaver DL} 1988.
Diffusion-collision model for the folding kinetics of myoglobin. \emph{Proteins
4}:211-227.

\noindent \textbf{Beck C, Siemens X, Weaver DL.} 2000. Non-Native Kinetic intermediates
and the Diffusion-Collision model of Protein Folding. \emph{In preparation}.

\noindent \textbf{Burton RE, Meyers JK,} Oas TG. 1998. Protein folding dynamics:
quantitative comparison between theory and experiment. \emph{Biochemistry 37}:5337-5343.

\noindent \textbf{Bryngelson JN, Onuchic ND, Socci ND, Wolynes PG.} 1995. Funnels,
pathways, and the energy landscape of protein folding: a synthesis. \emph{Proteins:
Struct Func and Gen 21}:need pages

\textbf{Chandler D}. 1987. Introduction to Modern Statistical Mechanics. Oxford
University Press.

\noindent \textbf{Chothia C.} 1974. Hydrophobic bonding and accessible surface
area in proteins. \emph{Nature 248}:338-339.

\noindent \textbf{Dinner AR, Lazaridis T, Karplus M.} 1999. Understanding \( \beta  \)-hairpin
formation. \emph{Proc Nat Acad Sci USA 96}:9068-9073.

\noindent \textbf{Fersht A.} Structure and mechanism in protein science. 158-159.
W.H. Freeman and Co. New York. 1999.

\noindent \textbf{Hanggi P, Talkner P, Borkovec M.} 1990. Reaction-rate theory
- 50 years after Kramers. \emph{Rev Mod Phys 62}:251-341.

\noindent \textbf{Karplus M, Weaver DL.} 1976. Protein folding dynamics. \emph{Nature
260}:404-406.

\noindent \textbf{Karplus M, Weaver DL.} 1979. Diffusion-collision model for
protein folding. \emph{Biopolymers 18}:1421-1437.

\noindent \textbf{Lacroix E, Viguera AR, Serrano L.} 1998. Elucidating the folding
problem of \( \alpha  \)-helices: Local motifs, long-range electrostatics,
ionic strength dependence, and prediction of of NMR parameters. \emph{J Mol
Biol 284}:173-191.

\noindent \textbf{Lee B, Richards FM.} 1971. The interpretation of protein structure:
Estimation of static accessibility. \emph{J Mol Biol 55}:379-400.

\noindent \textbf{Munoz V, Henry ER, Hofrichter J, Eaton WA.} 1998. A statistical
mechanical model for \( \beta  \)-hairpin kinetics. \emph{Proc Nat Acad Sci
USA 95}:5872-5879.

\noindent \textbf{Munoz V, Serrano L.} 1994a. Elucidating the folding problem
of helical peptides using empirical parameters. \emph{Nature: Struct Biol 1}:399-409.

\noindent \textbf{Munoz V, Serrano L.} 1994b. Elucidating the folding problem
of helical peptides using empirical parameters II: Helix macrodipole effects
and rational modification of the helical content of natural peptides. \emph{J
Mol Biol 245}:275-296.

\noindent \textbf{Munoz V, Serrano L.} 1994c. Elucidating the folding problem
of helical peptides using empirical parameters III: Temperature and pH dependence.
\emph{J Mol Biol 245}:297-308.

\noindent \textbf{Munoz V, Serrano L.} 1997. Development of the multiple sequence
approximation within the AGADIR model of \( \alpha  \)-helix formation. Comparison
with Zimm-Bragg and Lifson-Roig formalisms. \emph{Biopolymers 41}:495-509.

\noindent \textbf{Nagi AD, Anderson KS, Regan L} 1999. Using loop length variant
to dissect the folding pathway of a four-helix-bundle protein. \emph{J Mol Biol
286}:257-265.

\noindent \textbf{Pappu RV, Weaver DL.} 1997. The early folding kinetics of
apomyoglobin. \emph{Protein Science 7}:480-490.

\noindent \textbf{Regan L, Degrado WF.} 1988. Characterization of a helical
protein designed from first princliples. \emph{Science 241}:976-978.

\noindent \textbf{Szabo A, Schulten K, Schulten Z.} 1980. First passage time
approach to diffusion \emph{}controlled reactions. \emph{J Chem Phys 72}:4350.

\noindent \textbf{Van't Hoff JH.} 1884. Studies in chemical dynamics. London,
1896.

\noindent \textbf{Weaver DL.} 1979. Some exact results for one-dimensional diffusion
with absorption. \emph{Phys Rev B 20}:2558.

\noindent \textbf{Weiss GH.} 1967. First passage time problems is chemical physics.
\emph{Adv Chem Phys 13}:1-18.

\noindent \textbf{Zimm BH, Bragg JK.} 1959. Theory of phase transition between
helix and random coil in polypeptide chains. \emph{J Chem Phys 31}:526-535.

\end{document}